\documentclass[a4paper]{article}

\usepackage{INTERSPEECH2022}
\usepackage{float}
\usepackage[bottom]{footmisc}
\usepackage{cite}

\usepackage{booktabs, makecell, threeparttable, multirow}
\usepackage{array}
\usepackage{amssymb}
\usepackage{dsfont}
    \newcolumntype{P}[1]{>{\centering\arraybackslash}p{#1}}
    \newcolumntype{M}[1]{>{\centering\arraybackslash}m{#1}}
    
\newcommand{\dtoprule}{\specialrule{1pt}{0pt}{0.4pt}%
            \specialrule{0.3pt}{0pt}{\belowrulesep}%
            }

\title{Instance-Level Loss based Multiple-Instance Learning Framework for Acoustic Scene Classification}
\name{Won-Gook Choi$^1$, Joon-Hyuk Chang$^1$, Jae-Mo Yang$^2$, Han-Gil Moon$^2$}

\address{
  $^1$Department of Electronic Engineering, Hanyang University, Seoul, South Korea\\
    $^2$129, Samsung-ro, Yeongtong-gu, Suwon-si, Gyeonggi-do, 16677, South Korea}

\email{onlyworld94@naver.com, jchang@hanyang.ac.kr, jaemo81.yang@samsung.com, hangil.moon@samsung.com}

\begin{document}

\maketitle

\begin{abstract}
In the acoustic scene classification (ASC) task, an acoustic scene consists of diverse sounds and is inferred by identifying combinations of distinct attributes among them.
This study aims to extract and cluster these attributes effectively using an improved multiple-instance learning (MIL) framework for ASC.
MIL, known as a weakly supervised learning method, is a strategy for extracting an instance from a bundle of frames composing an input audio clip and inferring a scene corresponding to the input data using these unlabeled instances.
However, many studies pointed out an underestimation problem of MIL.
In this study, we develop a MIL framework more suitable for ASC systems by defining instance-level labels and loss to extract and cluster instances effectively.
Furthermore, we design a fully separated convolutional module, which is a lightweight neural network comprising pointwise, frequency-sided depthwise, and temporal-sided depthwise convolutional filters.
As a result, compared to vanilla MIL, the confidence and proportion of positive instances increase significantly, overcoming the underestimation problem and improving the classification accuracy up to 11\%.
The proposed system achieved a performance of 81.1\% and 72.3\% on the TAU urban acoustic scenes 2019 and 2020 mobile datasets with 139 K parameters, respectively.
Especially, it achieves the highest performance among the systems having under the 1 M parameters on the TAU urban acoustic scenes 2019 dataset.
\end{abstract}

\noindent\textbf{Keywords}: Acoustic scene classification, Multiple-instance learning, Weakly supervised learning

\section{Introduction}
Acoustic scene classification (ASC) is a task that aims to aware of acoustic environments by considering large contexts and combinations of sound events, mood, room size, culture, etc \cite{pham2019robust, bai19b_interspeech}.
It is widely accepted as a key functions of machine hearing \cite{lyon2017human} along with speech recognition, sound event detection (SED) and separation for acoustic environmental awareness.
Recently, ASC has been of special interest to audio artificial intelligence researchers using deep neural networks.
The achievement of an efficient ASC is also known to be the main challenge in the detection and classification of acoustic scenes and events (DCASE) community annually \cite{Martin2021, Heittola2020}.

ASC systems usually consider large or local contexts among complex acoustic data, unlike sound event detection and other audio tasks that usually treat target events or sources.
A particular sound event could represent an acoustic scene, but it is not unique as it could also appear in other scenes.
For example, a tram sound usually represents a tram acoustic scene (i.e., inside a moving tram); however, it may appear in a street pedestrian scene near a tramway.
In other cases, acoustic scenes comprise a combination of sound events, acoustic noises, and even echoes \cite{huwel2020hearing}.
There may even be no distinct events or attributes, such as silent situations or environments  \cite{pham2019robust, huwel2020hearing}.
It means that humans can recognize a park scene with sounds of moving water and birds chirping; however, if there are no events, they hard to distinguish whether it is a park or another outdoor scene despite it still being a park.

There have been many developments in ASC systems that treat the above problems with end-to-end learning methods since the adoption of deep neural networks.
In particular, convolutional neural networks (CNNs) significantly affect feature learning from handcrafted audio features, such as the log-mel spectrogram.
It is a common strategy for designing an ASC system to extract high-level feature maps consisting of scene activation scores from raw audio data and classify each scene using global average pooling (GAP) or fully connected layers through supervised learning methods \cite{wu2020time, kimqti}.

In contrast to these methods, another strategy aims to identify distinct sound events or frames from acoustic scenes without manual annotations.
\cite{song19b_interspeech} first used the multiple-instance learning (MIL) \cite{dietterich1997solving} in ASC, attempting to detect positive instances (i.e., frames containing distinct sound events or scene-characterizable frames) that can represent a scene.
MIL is a strategy that uses a relationship between a bag and instances to determine whether the bag (i.e., input data) is positive or negative.
Instances have hidden labels for identifying a bag as either positive or negative, and generally, the bag is considered positive if it contains at least one positive instance.
Considering that bags are labeled with the ground truths on audio clips but not on instances, the MIL-based ASC system is treated as weakly supervised learning owing to the decision method using unlabeled instances.
In other words, the bag is determined by the instances that are trained only with the bag-level label, without prior information about the instances.

An audio clip may have specific sounds representing a scene or only common sounds that can appear in multiple scenes.
For example, if babbling sounds with reverberation dominate the overall sound of an airport or a shopping mall, it makes it difficult to distinguish by hearings alone (we consider it an ambiguous sound or label).
Therefore, in many cases, a scene can be more precisely identified among the same category\footnote{The datasets for the DCASE challenge have defined ten acoustic scenes with three categories: airport, shopping mall and metro station into \emph{indoor}, street pedestrian, public square, street traffic, park into \emph{outdoor}, and metro, bus and tram into \emph{transportation}. In practice, many systems tend to misclassify within the same category.} if there are characteristic sounds within an audio clip.
In MIL-based ASC, a class of acoustic scenes is determined if the positive instances appear at least once.
Considering that MIL-based neural networks can accurately detect the positive and negative instances of an acoustic scene, they have great potential in ASC: acoustic scenes can be clearly distinguished even if they belong to the same category.
However, many studies pointed out the underestimation problem of a MIL-based neural network; in other words, the number of instances the system detects is too small  \cite{kolesnikov2016seed, kong2019sound, Yan:2017td, carbonneau2018multiple, wang2022frame}.
The most dominant positive instance determines the bag-level class, and the instances are trained on a bag-level label; therefore, the network predicts few positive instances, resulting in poor performance.

The problem arises because the bag-level loss is back-propagated sparsely to the instances.
This method naturally raises questions about whether the positive instances could be generated well and abundant.
We hypothesize that directly trained instances with some given information would help generate and cluster positive and negative instances.
Based on these questions and hypotheses, we reformulate the MIL framework for building ASC systems adopting instance-level loss.
\cite{wang2022frame} used a frame loss for weakly supervised SED based on the similar issue that the aggregated global loss had a high risk on frame-wise predictions.
The frame loss consists of an $\ell_2$-norm and a local smoothing term that keeps the inactive frames (i.e., frames without target events) at zero and maintains the regularity of adjacent active frames, respectively.

\begin{figure}[]
\centering
\includegraphics[width=\linewidth]{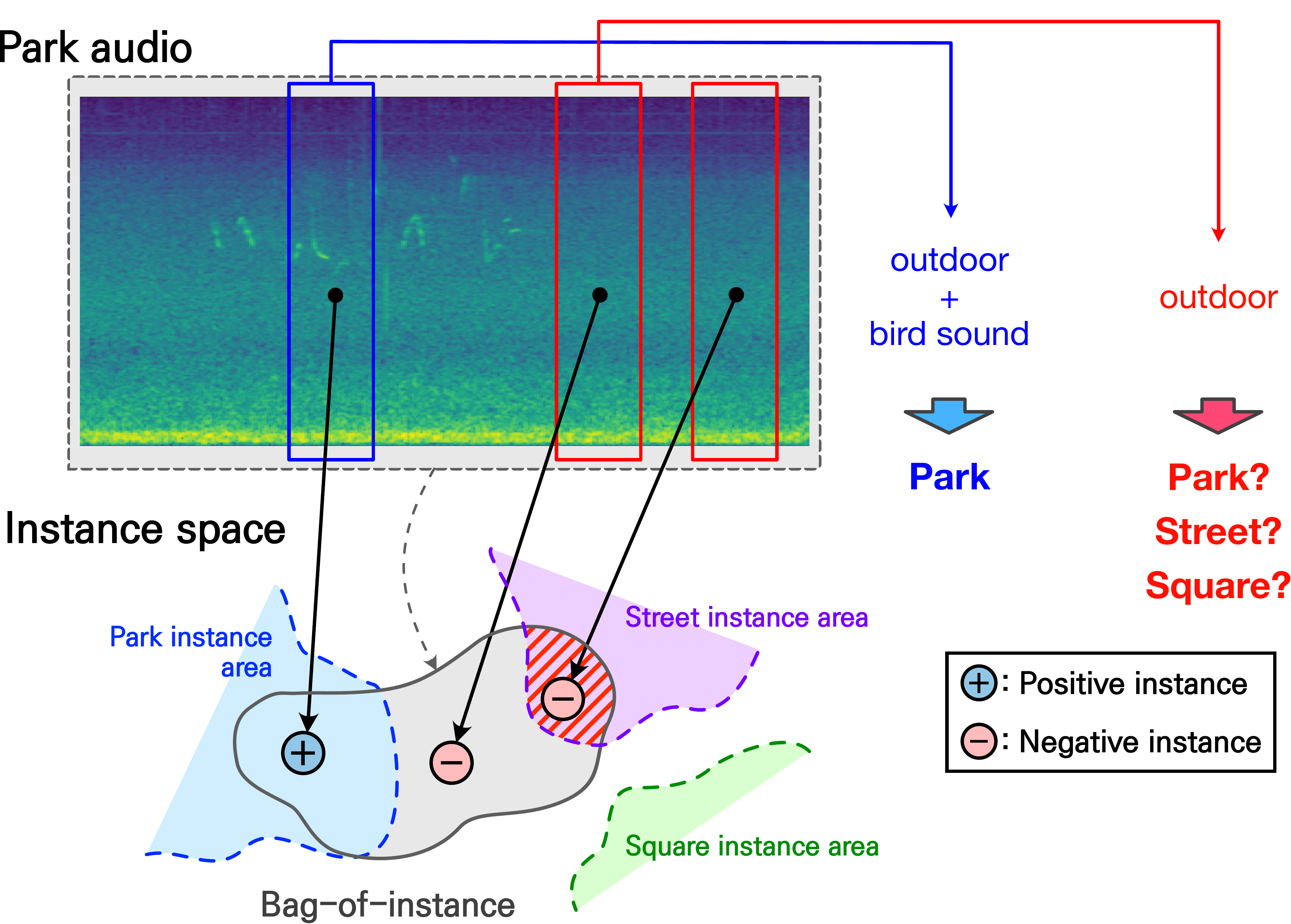}
\caption{\textbf{An example of the instances in an acoustic scene bag.} In the spectrogram of a park scene, the blue instance has a distinct attribute like the bird's sound, whereas the red instances has ambiguous properties. Ideally, there should be no intersection between the street or square instance area and the park bag. However, the intersection appears due to the label ambiguity, so the second negative instance falls into the street instance area.}
\label{fig:instance1}
\end{figure}

Unlike the SED task, where inactive frames could be distinguished from active frames, and the relationship among the adjacent frames could be helpful, the MIL-based ASC task suffers from a label ambiguity among the scenes (Fig.~\ref{fig:instance1}).
In this study, we introduce a strategy that defines and assigns the instance-level labels to calculate an instance-level loss.
The instance-level loss is calculated using the instance-level positive and negative labels generated by the instance-level confidences at each training step.
There are two effects of giving instances positive and negative labels.
First, each scene's positive and negative instances are well clustered in the instance-level space.
Second, because all instances are trained, positive instances are detected sufficiently.
Therefore, it can generate positive instances far more efficiently than systems without instance-level loss, thereby overcoming the underestimation problem.

Moreover, we propose an efficient CNN-based network for generating instances called a fully separated convolutional network, which consists only of axis-wise convolutional kernels rather than the box-shaped or spatial convolutional kernels.
When combined with the MIL framework, our network outperformed other CNNs, while dramatically reducing the number of parameters.

In the following section, we discuss the background for introducing instance-level loss along with the introduction of MIL.
The remainder of this paper discusses the details of instance-level loss and the instance generator, experiments, and results in Sections~\ref{sec:proposed}, ~\ref{sec:experiments}, and ~\ref{sec:Result}, respectively.

\section{Background: About Multiple-Instance Learning}
\label{sec:meaning}

\begin{figure}[t]
\centering
\includegraphics[width=\linewidth]{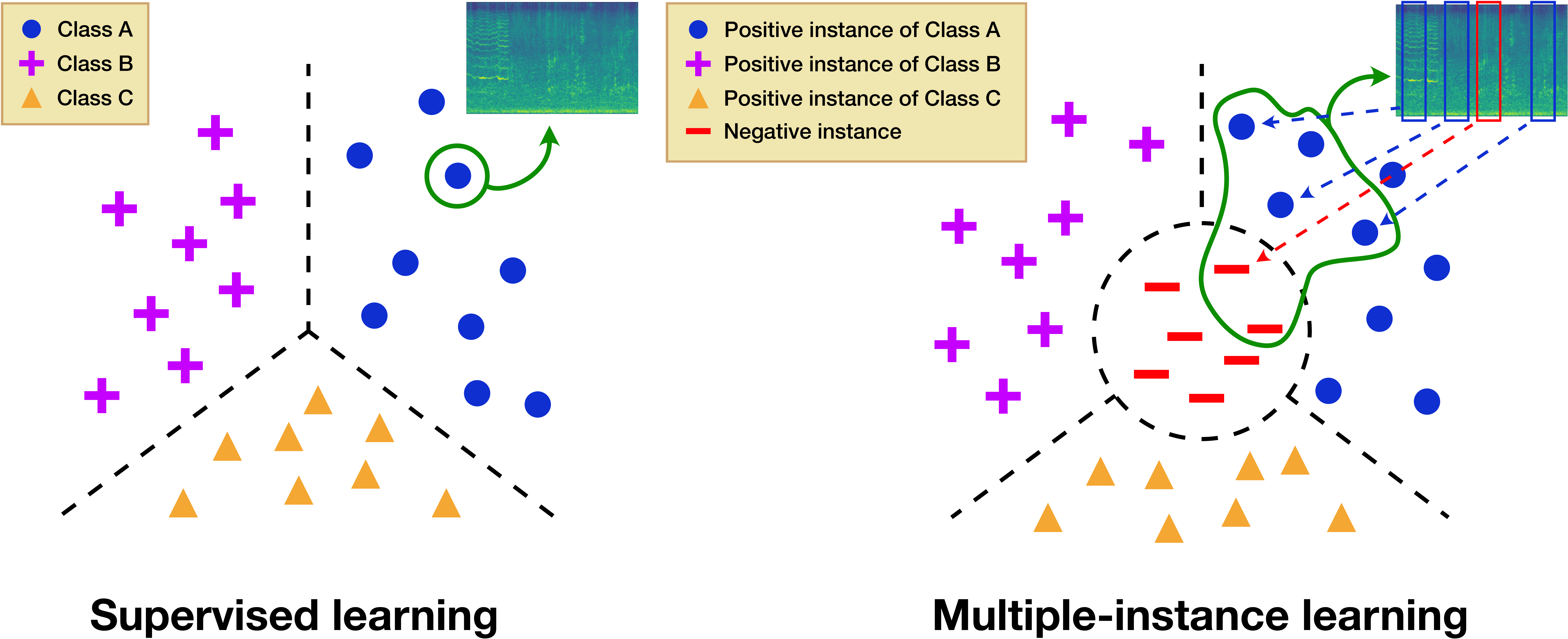}
\caption{\textbf{An illustration of decision boundaries on supervised learning and multiple-instance learning in audio domain.}}
\label{fig:concept}
\end{figure}

\begin{figure*}[ht]
\centering
\includegraphics[width=\linewidth]{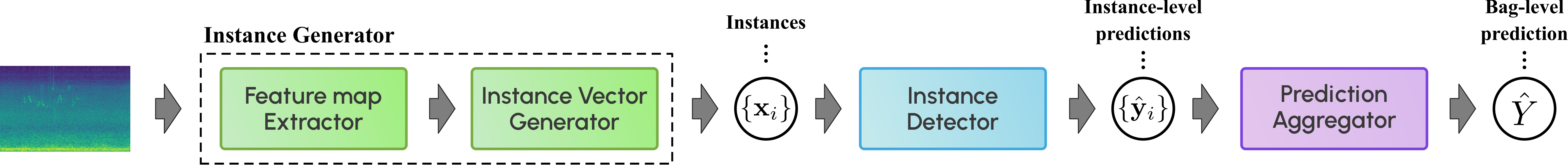}
\caption{\textbf{The overall structure of MIL based ASC system.}}
\label{fig:overall}
\end{figure*}
    
MIL is a widely used learning method  \cite{carbonneau2018multiple} that discovers representations of input data into a bag-of-instances (also called a bag), with the instances defined as representations of entities that constitute the bag.
Let \(X\) and \(f\) denote a handcrafted feature for the MIL-based system's input and an instance generator, respectively.
Then, the feature \(X\) is mapped to a bag-of-instances using the process \(f: X \rightarrow \mathcal{X} \triangleq \{\mathbf{x}_i\}_{i = 1}^n\), where \(\mathcal{X}\), \(\mathbf{x}_i\), and \(n\) denotes the bag-of-instances, the \(i\)-th instance belonging to the bag, and number of instances, respectively.

The label of \(X\) is inferred along with the bag composition with MIL assumptions, such as the standard MIL (SMI) assumption, count-based MIL (CMI) assumption, threshold-based MIL (TMI) assumption, etc  \cite{foulds2010review}, and the bag classifier is defined based on these assumptions.
The SMI assumption, which has been widely used in recent studies as well as in the past, states that the bag is positive if it contains at least one positive instance or negative otherwise.
In this case, the bag classifier \(g(\mathcal{X})\) is:
\begin{equation}
g(\mathcal{X}) = 
\begin{cases}
1, & \mbox{if} \ \exists \mathbf{x} \in \mathcal{X}: h(\mathbf{x}) = 1;\\
0, & \mbox{otherwise},
\end{cases}
\label{eq:SMI}
\end{equation}
where \(h(\mathbf{x})\) detects whether the instance \(\mathbf{x}\) is positive or negative.
If \(g(\cdot)\) is a multi-class classifier, (\ref{eq:SMI}) is turned to:
\begin{equation}
\begin{split}
\hat{\mathbf{y}}_i &= \mathbf{h}(\mathbf{x}_i), \ (0\leq i <N) \\
\hat{Y}_c &= \underset{i}{\mathrm{max}}\ \hat{y}_{ic}, \ (0\leq c <C) \\ 
g(\mathcal{X}) &= \mbox{argmax}_{c}\ \hat{Y}_c
\label{eq:SMI-multi}
\end{split}
\end{equation}
where \(N\) and \(C\) denote the number of instances and classes, respectively, \(\hat{y}_{ic}\) and \(\hat{Y}_c\) denote the c-th components of the instance-level and bag-level confidence vectors \(\hat{\mathbf{y}}_i,\ \hat{\mathbf{Y}} \in \mathbb{R}^c\), and \(\mathbf{h}\) detect whether the instance is positive or negative per class.
In other words, the classifier detects whether the bag is positive or negative for each class (the vector \(\hat{\mathbf{Y}}\) contains the best prediction scores in each class) and then determines the class with the highest confidence among the classes; the final class is determined by only the most dominant instance, regardless of the other instances.
A definition of positive instance depends on an MIL assumption: in the SMI assumption, an instance is positive if the argument of the maxima (argmax) of the instance-level confidence vector for the number of classes is the same as that of the bag-level.

Instances are defined as bag components, so the bag-level space results in a set of subsets of the instance-level feature space \cite{doran2016multiple}.
The bag-level probability distribution is calculated using the set of the instance-level probability distribution.
Therefore, it is essential to train the instances well because the bag is a group of instances, and the instances are involved in determining the bag's label.

According to the previous studies in various fields, each field defines an instance differently.
For example, in the recent audio field \cite{wang2018polyphonic, song19b_interspeech, hong20_interspeech, wang2022frame}, an instance is generated from a bundle of frames in a spectrogram.
In an image classification task, an instance is extracted from a small patch of an image \cite{carbonneau2018multiple}.
For text data, an instance is extracted from a sentence composing a document \cite{lutz2018sentence}.
Considering the studies above, an instance is an intuitive representation generated by a sub-concept of the input data domain, not just a totally abstract latent vector.
Thus, we can know which parts (i.e., instances) contribute to making the data positive.

\subsection{MIL based ASC}
MIL-based ASC was first proposed in \cite{song19b_interspeech} to identify distinct sounds for classifying the scenes, which was inspired by a psychological study in \cite{peltonen2001recognition}.
They aimed to detect instances that contained the scene-specific sound events, and these were defined as positive instances.

The overall architecture is illustrated in Fig.~\ref{fig:overall}.
An instance generator consists of a feature map extractor and an instance vector generator: extracts high-level feature maps from a log-mel spectrogram, and generates instance vectors from the feature maps.
For examples, the shape of the feature map extractor's output is a tensor that shapes (\emph{channel, frequency, frame}).
The instance vector generator then aggregates the tensor using a convolutional layer with a full-size (i.e., fully connected) kernel along the frequency axis (i.e., kernel size: \((freq. \times 1)\)).
Consequently, the instances are extracted from a bundle of spectrogram frames, and their dimension is the same as the number of CNN's channels in the feature map extractor.

The instance detector calculates whether each instance is positive or negative using softmax or sigmoid activation.
Subsequently, the prediction aggregator uses a max-pooling operation to combine the instances into a bag.
Finally, for the final decision, bag-level prediction is used.

The loss is introduced by the mean of the \emph{weighted binary cross entropies} (wBCEs), which is used to calculate the mean of the BCEs for each class by considering the imbalance of positive and negative classes.
Eq. \eqref{eq2} shows the mean of the wBCEs (\(\overline{wBCEs}\)), where \(C\), and \(\alpha\) are the number of the classes, and the class imbalance factor, respectively, and \(\alpha\) was set to \(C-1\).
If \(\alpha\) is one, wBCE is the same as BCE, which ignores the class imbalance.
    \begin{equation}
    \begin{split}
      \overline{wBCEs}_{bag} &= 
      -\frac{1}{C}\sum_{c = 1}^{C}
        (
        \alpha \cdot Y_{c} \textrm{ln}{\hat{Y}_c} +
        (1 - Y_{c}) \textrm{ln}{(1 - \hat{Y}_c)}
        )
      \label{eq2}
    \end{split}
    \end{equation}

\subsection{Label ambiguity in MIL}
\label{sec:prop2}
Label ambiguity is an MIL problems caused by the different label spaces at the instance-level and bag-level  \cite{carbonneau2018multiple}.
Ideally, a positive instance from one scene cannot appear in the other scenes \cite{song19b_interspeech}.
For example, in Fig.~\ref{fig:instance1}, the spectrogram is mapped to the park scene bag, and the stippled areas are groups of distinct instances (i.e., positive instances) that can represent enough each predefined scene (i.e., park, street, and square in the figure).
The red-shaded area indicates where the bag and other groups of positive instances intersect; ideally, it should not appear.
However, the negative instances extracted from the target scene could fall into a positive area of the other scenes owing to the label ambiguity.

In this study, we consider the label ambiguity to be natural rather than forcing a decision boundary to pass ambiguous instances because there are common characteristics among the scenes, such as the ambiguity between the serene situations of a park and a public square.
One way to design a system reflecting label ambiguity is to control the objective function: either CE or BCE.
Most of the ASC tasks used categorical cross-entropy (CE) for the objective function whereas we adopt BCE, as in \cite{song19b_interspeech}.
Note that in this paper, what we mention training the network using BCE means that wBCE is used in practice.
The experimental results and effects according to the objective functions will be discussed in Section~\ref{sec:objective}.

        \begin{figure*}[ht]
         \centering
         \includegraphics[width=\linewidth]{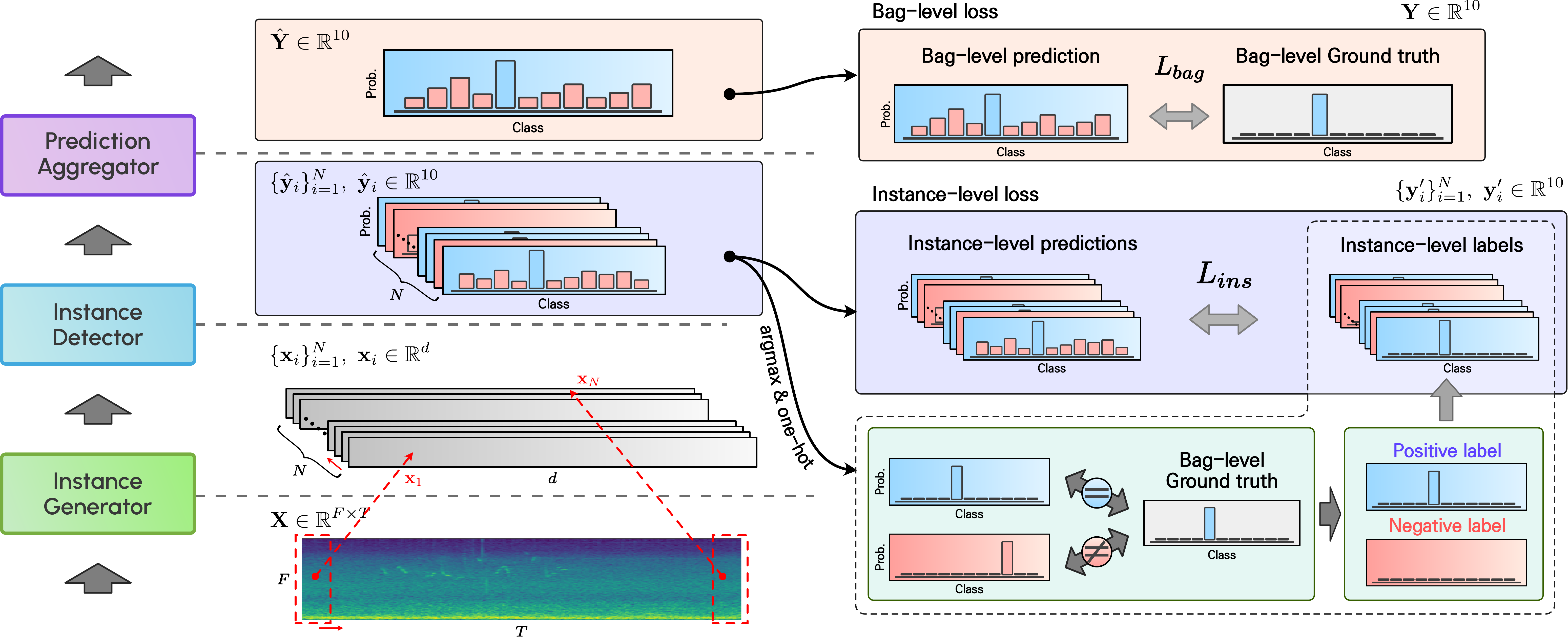}
         \caption{\textbf{Schematic diagram of MIL with instance-level loss.} \(\textbf{X}\), \(\textbf{x}_i\), \(N\), and \(d\) denote an input spectrogram, an instance, the number of instances, and the dimension of instances, respectively. And \(\hat{\textbf{Y}}\), \(\hat{\textbf{y}}\), \(\textbf{Y}\), and \(\textbf{Y}^{\prime}_{i}\) denote the inference of the bag and instance, and the bag-level and instance-level label. \(T\) is mapped to \(N\) by the instance generator.}
         \label{fig:ins_loss}
        \end{figure*}

\section{Proposed Method}
\label{sec:proposed}

\subsection{Instance-level loss}
\label{sec:instance_label}
We set an instance-level loss, which leads to training the instances directly and helps cluster positive and negative instances well, unlike the previous study of an MIL-based ASC \cite{song19b_interspeech}, which had only considered a bag-level loss.
In contrast to the frame loss \cite{wang2022frame} that uses $\ell_2$-norm and the relationship between instances, the proposed instance-level loss is based on a temporarily defined label.
Therefore, we define the instance-level label to derive the instance-level loss.
Because the instances lack labels, we assign \emph{positive or negative labels} to the instances using both bag-level ground truths and instance-level confidences in each training step.
By doing this, we could avoid manual labeling and expect that instances are well clustered to be positive or negative in the instance space.

Eq. \eqref{eq4} and Fig.~\ref{fig:ins_loss} show details defining instance-level label using bag-level ground truths and instance predictions.
We assign a positive label if the class with the greatest confidence at each instance is the same as the bag-level ground truth or otherwise, we assign a negative label.
A positive label is defined as a one-hot vector (\(\mathbf{e}^{(c)} \in \mathbb{R}^{10}\)), while a negative label is defined as a zero vector.
Similar to the bag-level loss, we use the mean of the wBCEs for instance-level loss, which allowed the model to cluster the distinct instances and negatives.
            \begin{equation}
            \begin{split}
              L_{ins} &= 
              -{\frac{1}{NC}}\sum_{i = 1}^{N}\sum_{c = 1}^{C}
                (
                \alpha \cdot y_{ic}^{\prime} \, \textrm{ln}\, {\hat{y}_{ic}} +
                (1 - y_{ic}^{\prime}) \textrm{ln}{(1 - \hat{y}_{ic}}
                )) \\
              \mathbf{y}^{\prime}_{i} &= 
                  \begin{cases}
                    \mathbf{e}^{(t)}, & \mbox{if argmax}_{c} \ \hat{y}_{ic} = t, t\mbox{: ground truth}\\
                    \mathbf{0}, & \mbox{otherwise}
                  \end{cases}  
              \label{eq4}
            \end{split}
            \end{equation}
As a result, the total loss can be expressed as follows:
            \begin{equation}
          L_{total} = L_{bag} + L_{ins}.
          \label{eq1}
        \end{equation}
        
All parameters are set randomly at the beginning of the training, so instance-level labels are also set randomly, whether the instances are distinctive or not.
Fortunately, because the bag-level loss is calculated with the ground truth, it correctly guides the instance-level label to the right way as the training step progresses.
Otherwise, the instance-level labels would fall into the far wrong label space.

Similar to other MIL approaches, this above strategy maintains the concept of weakly supervised learning.
As a side note, we also attempted the MIL framework-based supervised learning, which gives an instance a \emph{bag-level ground truth} rather than a positive or negative label.
Generally, the instances do not have labels; however, we assume that the label is known.
Therefore, setting the bag-level ground truth as an instance-level label is an approach that combines the MIL framework with supervised learning, rather than the MIL itself.
In fact, a difference between MIL and traditional supervised learning is that the number of instances is one or more: in terms of MIL, the traditional supervised learning method is treated as single-instance single-label learning, in which the instance and bag are matched one-to-one  \cite{zhou2012multi}.
In other words, setting the bag-level ground truth as an instance-level label is an approach that combines the MIL framework with supervised learning rather than the MIL itself.
In terms of traditional supervised learning, it is similar to \emph{ensemble system} that breaks an audio clip into smaller segments for inference and chooses the most confident value from the results of each segment; in this case, bags and instances share the same label space because the negative label is not defined at the instance-level.
Note that the fact does not change that the most influential instance determines the scene, which has the same decision-making as the MIL's standard assumption.
The results are presented in the Section~\ref{sec:PNvsGT}

    \subsection{Fully separated convolutional network}

        \begin{figure}[ht]
          \centering
          \includegraphics[width=\linewidth]{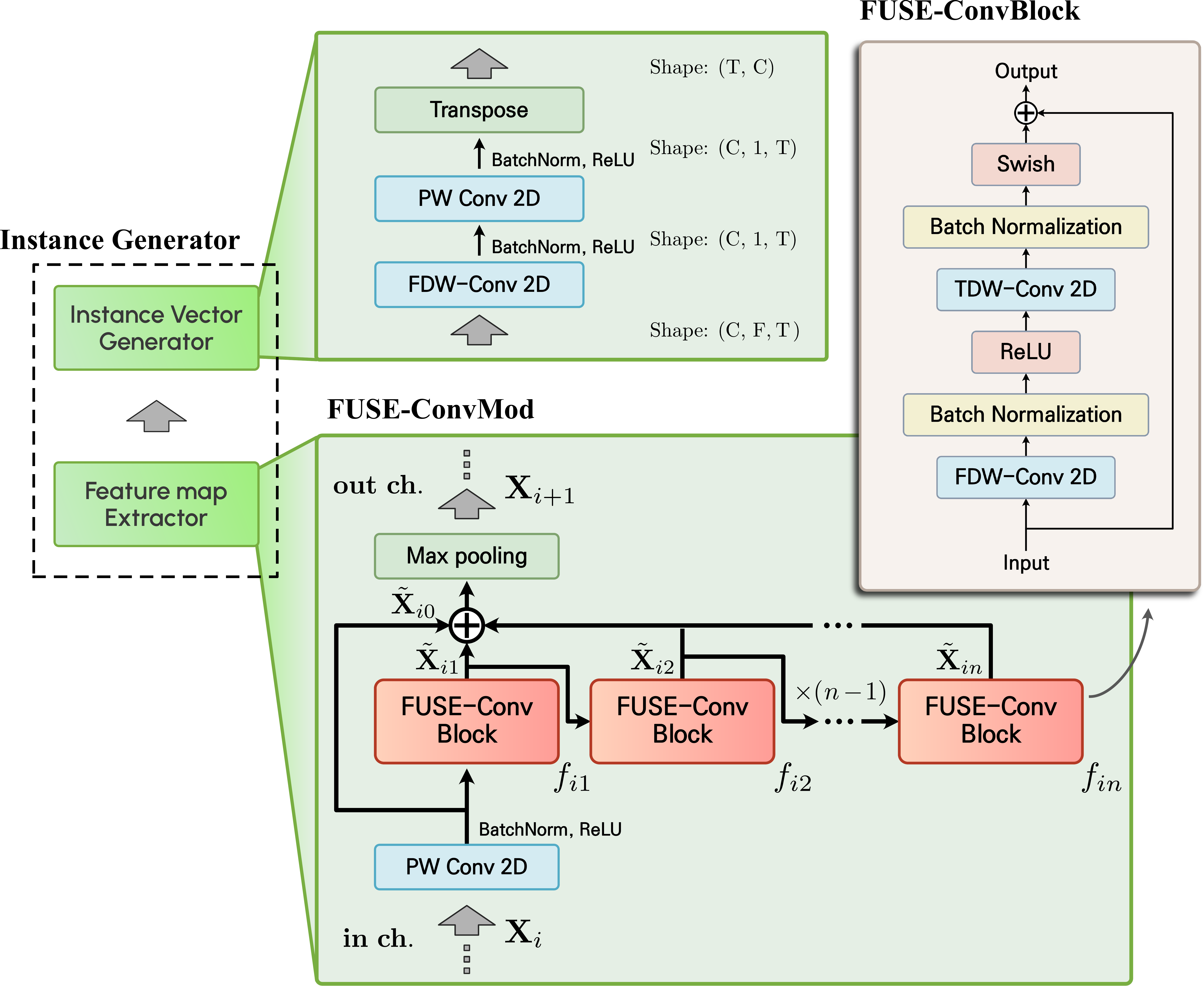}
          \caption{\textbf{Architecture of the instance generator.} It consists of the feature map extractor which is made up of several FUSE-ConvMods, and the instance vector generator composed full-sized FDW-Convlayer and PW Conv2d. In the FUSE-ConvMod, \(n\) denotes the number of FUSE-ConvBlocks.}
          \label{fig:instance_generator}
        \end{figure}

        Instead of using the VGG (visual geometry group)-like feature extractor in \cite{song19b_interspeech}, we apply spatially separable convolutional layers \cite{szegedy2016rethinking} with pointwise (PW) and depthwise (DW) convolution layers \cite{howard2017mobilenets}.
        
        In terms of the complexity, the number of parameters of a standard convolutional layer is:
            \begin{equation}
            \label{comp1}
                C_{in} \cdot C_{out} \cdot K_{H} \cdot K_{W},
            \end{equation}
        where the complexity depends on the multiplication of the input and output channel sizes \(C_{in}\) and \(C_{out}\), and the height and width of the kernel \(K_{H}\) and \(K_{W}\).
        Depthwise separable convolution~\cite{howard2017mobilenets} is an efficient version of using the channel-wise convolutional filter while maintaining a receptive field.
	    It consists of two types of convolutional layers: DW and PW convolutional layers.
        The complexity of a depthwise separable convolutional layer is calculated as follows:
            \begin{equation}
            \label{comp2}
                \underbrace{C_{in} \cdot C_{out}}_{Pointwise}
                + \underbrace{C_{out} \cdot K_{H} \cdot K_{W}}_{Depthwise} \\
                = C_{out} ( C_{in} + K_{H} \cdot K_{W} ).
            \end{equation}
        The spatially separable convolutional layer contains totally factorized convolutional filters.
        It replaces the \(3 \times 3\) filter with a combination of \(3 \times 1\) and \(1 \times 3\) filters.
        The complexity of a spatially separable convolutional layer is expressed as follows:
            \begin{equation}
            \begin{split}
            \label{comp3}
                &\underbrace{C_{in} \cdot C_{out}}_{Pointwise}
                + \underbrace{C_{out} \cdot K_{H}}_{(K_H \times 1) \ Depthwise}
                + \underbrace{C_{out} \cdot K_{W}}_{(1 \times K_W) \ Depthwise} \\ \\
                &= C_{out} ( C_{in} + K_{H} + K_{W} ).
            \end{split}
            \end{equation}
	Note that the output channel size \(C_{out}\) is equal to the number of convolutional filters; therefore, the complexity of the spatially separable convolutional layer is calculated soley by the addition operation instead of multiplication.
	In other words, the model complexity is kept small although the receptive fields (i.e., filter size) and the number of channels grow larger.
        
        In contrast to image data that are worth spatial information, a spectrogram shows different properties along the frequency and time axes.
        Therefore, the kernels from the spatially separable convolution become the frequency-side and temporal-side kernels.
        According to previous studies with the spatially separable convolution in audio fields, the keyword spotting \cite{kim2021broadcasted} and ASC \cite{kimqti} tasks have been shown to be beneficial for modeling fewer parameters as well as for performance improvements.

	Based on previous studies, we designed a new CNN for the feature map extractor: a \emph{fully separated convolutional network} (FUSE, Figs. \ref{fig:instance_generator}).
	FUSE consists of several \emph{fully separated convolutional modules} (FUSE-ConvMods) and a FUSE-ConvMod consisting of several \emph{fully separated convolutional blocks} (FUSE-ConvBlocks), which have harmonies of PW, frequency-side DW (FDW), and temporal-side DW (TDW) convolutional layers.
        An input feature passes through a PW convolutional layer followed by several FUSE-ConvBlocks, and is pooled through a max-pooling layer.
        Mathematically, the output employed by the \(i\)-th FUSE-ConvModule is:
        \begin{equation}
        \begin{split}
        \label{FUSE-ConvModule}
            \tilde{\mathbf{X}}_{i0} &= \mbox{PW-Conv2d}(\mathbf{X}_{i})  \\
            \tilde{\mathbf{X}}_{ik} &= f_{ik}(\tilde{\mathbf{X}}_{ik-1}) \ (k = 1...n)  \\ 
            \mathbf{X}_{i+1} &= \mbox{MaxPooling}( \sum_{k = 0}^{n} \tilde{\mathbf{X}}_{ik}),
        \end{split}
        \end{equation}
        where \(\mathbf{X}_{i},\ N\), and \(f_{ik}\) denote the input of the \(i\)-th FUSE-ConvMod, the number of the FUSE-ConvBlock and \(k\)-th block in the \(i\)-th FUSE-ConvMod, respectively, as shown in Fig.~\ref{fig:instance_generator}.
        As an exception, we do not carry out pooling for the last FUSE-ConvMod.

\section{Experiments}
\label{sec:experiments}
\subsection{Datasets}
Some of the most popular datasets for the ASC are the TAU urban acoustic scenes datasets \cite{heittola_toni_2019_2589280, heittola_toni_2019_2589332, heittola_toni_2019_2591503, heittola_toni_2020_3670185, heittola_toni_2020_3819968}.
For years, the TAU urban acoustic scenes datasets have been used for the DCASE challenges for various purposes; therefore, there are many versions of the TAU urban acoustic scenes datasets.
To evaluate the performance of the proposed method, we experimented with the \emph{TAU urban acoustic scenes 2019 dataset} (\emph{TAU 2019})  \cite{heittola_toni_2019_2589280} and the \emph{TAU urban acoustic scenes 2020 mobile dataset} (\emph{TAU 2020m}) \cite{heittola_toni_2020_3819968}.
Furthermore, we conducted a simple experiment to compare the proposed method, the superiority of the instance-level loss, to the study \cite{song19b_interspeech} that first proposed the MIL-based ASC system.

\emph{\textbf{TAU 2019}} was first released in the DCASE 2019 challenge task 1-A, which aimed to classify ten acoustic scenes collected with a single device in 12 European cities.
It was recorded by using a 48 kHz sampling rate with stereo channels.
The 40 h of audio recordings consist of 14,400 segments in ten different acoustic scenes, and each segment has 10 s long.
There are 9,185 training segments and 4,185 validation segments.

\begin{table}[ht]
\caption{\textbf{Details of the instance generator.} A row shows a stage (e.g. feature map extractor or instance vector generator). \(\textbf{n}\), \(\textbf{c}\), and \(\textbf{p}\) denotes the number of blocks repeated in the FUSE-ConvMod, the output channel of each module, and pooling or not at the end of the FUSE-ConvMod, respectively. Batch size is not included in the input shape.}
\label{tab:instance_generator}
\centering
{\footnotesize
\begin{tabular}{r|ccccc}
\toprule
\textbf{Stage} & \textbf{Input shape} & \textbf{Network} & \textbf{n} & \textbf{c} & \textbf{p} \\ \midrule
Feature & 1, 256, T & FUSE-ConvMod & 3 & 32 & O \\
map & 32, 128, T/2 & FUSE-ConvMod & 3 & 64 & O \\
Extractor & 64, 64, T/4 & FUSE-ConvMod & 3 & 128 & O \\
& 128, 32, T/8 & FUSE-ConvMod & 3 & 256 & X \\ \midrule
Instance & 256, 32, T/8 & FDW-Conv2D & - & 256 & - \\ 
vector & 256, 1, T/8 & PW-Conv2D & - & 256 & - \\
Generator & 256, 1, T/8 & Reshape & - & - & - \\
& 256, T/8 & Transpose & - & - & - \\
\bottomrule
\end{tabular}
}
\end{table}

\emph{\textbf{TAU 2020m}} differs from the TAU 2019; this dataset was used to study device generalization for the minor and unseen recording devices in the DCASE 2020 and 2021 challenge task 1-A.
Additional recordings were provided, including those from TAU 2019.
Three devices were used to collect real recordings: one was the device used in TAU 2019 (called device A), and the other devices were smartphones (called device B and C).
Furthermore, six types of simulated recordings were generated by applying filters with six different frequency responses to the real recordings from device A (called device S1-S6).
Each has a 3 h audio recording and simulation from device B to S6.
All recordings were constructed with a 44.1 kHz sampling rate with the mono channel and consisted of 23040 segments.
The validation data for each device was set to 330 segments for cross-validation, and the remaining segments were used for training.
Note that among the remaining devices, S4 to S6 were not used in the network's training so, were treated as unseen devices to generalize the model against mismatched and unseen devices, as stated in the task.

\emph{\textbf{TUT 2018m}} \cite{heittola_toni_2018_1228235} is a minor version of TAU 2020m that aimed to generalize the device properties of mismatched devices.
It consists of devices A, B, and C, each with 24 h, 2 h, and 2 h durations.
Similarly, the recordings of devices A, B, and C were split into 6122, 540, and 540 segments for training, and 2518, 180, and 180 segments for validation.

\subsection{Training setup}
Before the feature extraction, each segment was downmixed to 16 kHz.
Subsequently, log-mel spectrograms were extracted for input features with a window size of 128 ms, hop size of 32 ms, and 256 mel-filter banks.
We optimized with the stochastic gradient descent optimizer and learning rate scheduler using cosine annealing  \cite{LoshchilovH17} with warming up the first five epochs for 100 epochs.
The initial learning rate, weight decay, and batch size were set as 0.06, 0.001, and 48, respectively.
The details of the instance generator are presented in Table~\ref{tab:instance_generator}.
Additionally, we used a sigmoid function following the \((256 \times 10 )\) linear layer for the instance detector and a max-pooling layer for the prediction aggregator.

\section{Results and Discussions}
\label{sec:Result}
We formed four main comparison groups to demonstrate the superiority of the proposed instance-level loss and FUSE.
\begin{itemize}
\item The vanilla MIL that does not employ the instance-level loss (MIL w/o IL, Section~\ref{sec:wnw/o}).
\item The conventional supervised learning method using a classifier with GAP (SupL w/ GAP, Section~\ref{sec:tradSupL}).
\item CE loss for the objective function to instance-level loss for regularizing the label ambiguity (MIL w/ PNL CE, Section~\ref{sec:objective}).
\item The feature map extractor in  \cite{song19b_interspeech} (Section~\ref{sec:resultFUSE}).
\end{itemize}

{
\centering
\footnotesize
\begin{tabular}{cc} \midrule[1pt]
\multirow{1}{*}{\makecell{Abbreviation}} & \multirow{1}{*}{\makecell{Description}}  \\ \midrule[1pt]
\multirow{1}{*}{\makecell{MIL w/o IL}} & \multirow{1}{*}{\makecell{MIL without instance-level loss (vanilla MIL)}} \\ \cmidrule{1-2}
\multirow{2}{*}{\makecell{SupL w/ GAP}} & \multirow{1}{*}{\makecell{Conventional supervised learning method \\ using a classifier with GAP}} \\
~ & ~ \\  \cmidrule{1-2}
\multirow{2}{*}{\makecell{MIL w/ PNL BCE}} & \multirow{1}{*}{\makecell{MIL assigned positive or negative labels \\ for instance-level loss using BCE}} \\
~ & ~ \\  \cmidrule{1-2}
\multirow{2}{*}{\makecell{MIL w/ PNL CE}} & \multirow{1}{*}{\makecell{MIL assigned positive or negative labels \\ for instance-level loss using CE}} \\
~ & ~ \\  \cmidrule{1-2}
\multirow{2}{*}{\makecell{MIL w/ GT BCE}} & \multirow{1}{*}{\makecell{MIL assigned bag-level ground truth \\ for instance-level loss using BCE}} \\
~ & ~ \\  \midrule[1pt]
\end{tabular}
}

We mainly used the SMI assumption for the experiments, but CMI and TMI assumptions were also discussed in Section~\ref{sec:assumption}.
Each experiment was repeated with five random seeds.

\subsection{Effects of instance-level loss}
\label{sec:ILLoss}
The effect of instance-level loss was significant in the MIL-based ASC.
As shown in Table~\ref{tab:experiments}, the classification accuracy of the MIL-based ASC system in  \cite{song19b_interspeech} was improved by approximately 5\% after adopting instance-level loss.
Furthermore, the MIL-based systems with both VGG-like and FUSE extractors on the TAU 2019 and 2020m also improved significantly.
Additionally, we experimented with various CNNs, frequently used and shown good performance in ASC tasks, as feature map extractors on the TAU 2019 and 2020m to show the superiority of instance-level loss, and the results are shown in Table~\ref{tb:exp_result}.
In most cases, the conventional supervised learning method outperformed the vanilla MIL, but the proposed method outperformed the supervised learning method.

Additionally, we compared our proposed method to the frame loss proposed in \cite{wang2022frame} (Table~\ref{tab:frameloss}).
As expected, frame loss was effective for SED but not for ASC.
SED aims to recognize the patterns of target sounds, and these active frames are distinguished clearly compared to inactive frames.
Therefore, the objective functions that minimize the $\ell_2$-norms of inactive frames and variance of active frames were appropriate for training the SED system.
However, the acoustic scene has complex contexts; therefore, the sounds could be part of the same scene, even if their statistical patterns differ.

\begin{figure*}[ht]
\centering
\includegraphics[width=\linewidth]{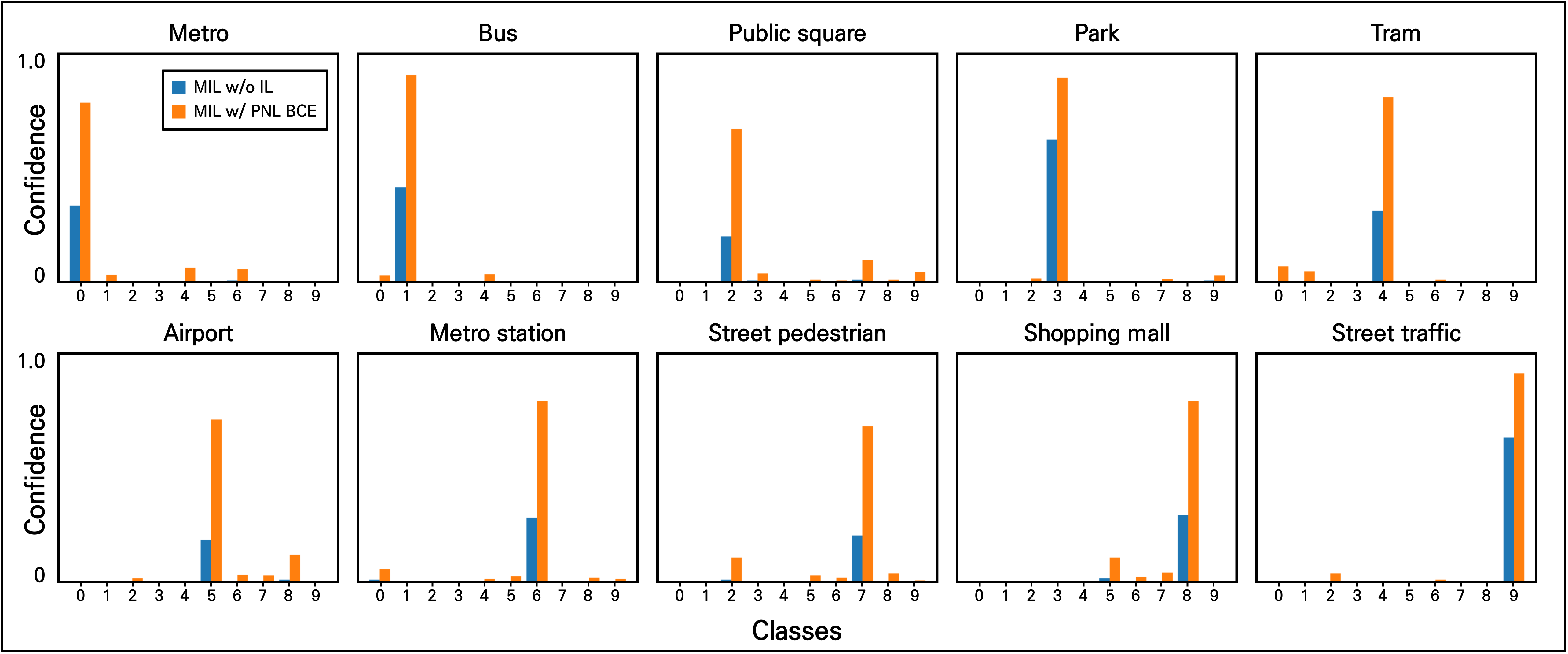}
\caption{\textbf{Averages on instance-level confidence for each class.} (0: metro, 1: bus, 2: public square, 3: park, 4: tram, 5: airport, 6: metro station, 7: street pedestrian, 8: shopping mall, 9: street traffic)}
\label{fig:TPR}
\end{figure*}

\begin{figure}[ht]
\centering
\includegraphics[width=\linewidth]{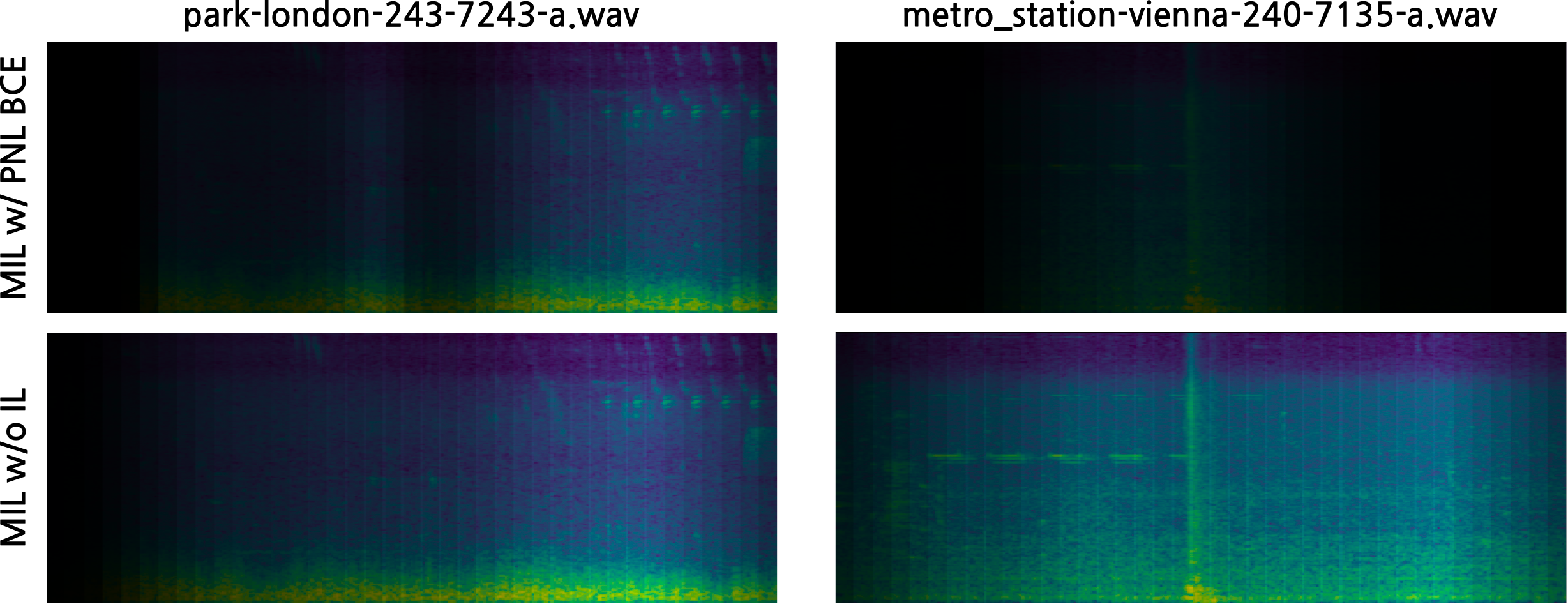}
\caption{\textbf{Masked spectrograms based on the confidence of instances.} The park audio is an well-estimated sample whereas, the metro station audio is an underestimated sample for the vanilla system. The brighter, the more distinct instance it was.}
\label{fig:spec}
\end{figure}

\begin{table}[h]
\centering
\caption{\textbf{Comparison of the classification performance to the MIL based ASC system proposed in \cite{song19b_interspeech}.}(*: the performance reported in \cite{song19b_interspeech}, w/ IL means w/ PNL BCE.)}
{
\footnotesize
\begin{tabular}{cccccc} \dtoprule

\multirow{2}{*}{\makecell{Dset}} & \multirow{2}{*}{\makecell{Feature map\\extractor}} & \multirow{2}{*}{\makecell{Mel\\bins}} & \multirow{2}{*}{Params} &  \multicolumn{2}{c}{Acc(\%)} \\ \cmidrule(lr){5-6}
~ & ~ &  ~ & ~  & w/o IL & w/ IL \\ \midrule[1pt]

\multirow{2}{*}{\makecell{TUT\\2018m}} & \multirow{2}{*}{\makecell{VGG-like}} & \multirow{2}{*}{\makecell{40}} & \multirow{2}{*}{\makecell{454K}} & \multirow{2}{*}{*64.20} & \multirow{2}{*}{\textbf{69.57}} 
\\ \\ \midrule[1pt]

\multirow{2}{*}{\makecell{TAU\\2019}} & \multirow{1}{*}{\makecell{VGG-like}} & \multirow{1}{*}{\makecell{256}} & \multirow{1}{*}{\makecell{1.34M}} & \multirow{1}{*}{66.35} & \multirow{1}{*}{\textbf{77.17}} \\ \cmidrule{2-6}
~ & \multirow{1}{*}{\makecell{FUSE}} & \multirow{1}{*}{\makecell{256}} & \multirow{1}{*}{\makecell{139K}} & \multirow{1}{*}{72.22} & \multirow{1}{*}{\textbf{79.13}} \\ \midrule[1pt]

\multirow{2}{*}{\makecell{TAU\\2020m}} & \multirow{1}{*}{\makecell{VGG-like}} & \multirow{1}{*}{\makecell{256}} & \multirow{1}{*}{\makecell{1.34M}} & \multirow{1}{*}{58.17} & \multirow{1}{*}{\textbf{66.31}} \\ \cmidrule{2-6}
~ & \multirow{1}{*}{\makecell{FUSE}} & \multirow{1}{*}{\makecell{256}} & \multirow{1}{*}{\makecell{139K}} & \multirow{1}{*}{60.13} & \multirow{1}{*}{\textbf{66.75}} \\ \bottomrule[1.5pt]

\end{tabular}
}
\label{tab:experiments}
\end{table}

\subsubsection{With and without the instance-level loss}
\label{sec:wnw/o}

\begin{table}[ht]
\centering
\caption{\textbf{Experimental results according to the learning strategies for various CNNs.} Residual network (ResNet): \cite{he2016deep}, the number of blocks, stacks, and filters are \{3, 3, 3, 1\}, \{2, 2, 2, 2\}, and \{8, 16, 32, 64\}, respectively. Broadcasted ResNet (BC-ResNet): \cite{kimqti})}
{
\footnotesize
\begin{tabular}{cccc} \dtoprule

\multirow{2}{*}{\makecell{Extractor}} & \multirow{2}{*}{\makecell{Learning\\strategy}} & \multicolumn{2}{c}{Acc(\%)} \\ \cmidrule(lr){3-4}
~ & ~ & TAU 2019 & TAU 2020m \\ \midrule[1pt]

\multirow{4}{*}{\makecell{VGG-like}} & \multirow{1}{*}{\makecell{SupL w/ GAP}} & \makecell{75.17} & 65.17 \\ \cmidrule{2-4}
~ & \makecell{MIL w/o IL} & \makecell{66.35} & 58.17 \\ \cmidrule{2-4}
~ & \makecell{MIL w/ PNL BCE} & \textbf{77.17} & \textbf{66.31} \\ \cmidrule{1-4}

\multirow{4}{*}{\makecell{ResNet}} & \multirow{1}{*}{\makecell{SupL w/ GAP}} & \makecell{71.67} & 63.14 \\ \cmidrule{2-4}
~ & \makecell{MIL w/o IL} & \makecell{71.46} & 58.59 \\ \cmidrule{2-4}
~ & \makecell{MIL w/ PNL BCE} & \textbf{75.30} & \textbf{64.05} \\ \cmidrule{1-4}

\multirow{4}{*}{\makecell{BC-ResNet}} & \multirow{1}{*}{\makecell{SupL w/ GAP}} & \makecell{77.02} & 65.36 \\ \cmidrule{2-4}
~ & \makecell{MIL w/o IL} & \makecell{74.77} & 63.46 \\ \cmidrule{2-4}
~ & \makecell{MIL w/ PNL BCE} & \textbf{77.39} & \textbf{66.29} \\ \cmidrule{1-4}

\multirow{4}{*}{\makecell{FUSE}} & \multirow{1}{*}{\makecell{SupL w/ GAP}} & \makecell{71.61} & 64.73 \\ \cmidrule{2-4}
~ & \makecell{MIL w/o IL} & \makecell{72.22} & 60.13 \\ \cmidrule{2-4}
~ & \makecell{MIL w/ PNL BCE} & \textbf{79.13} & \textbf{66.75} \\ \midrule[1pt]

\end{tabular}
}
\label{tb:exp_result}
\end{table}

\begin{figure}[h]
\centering
\includegraphics[width=\linewidth]{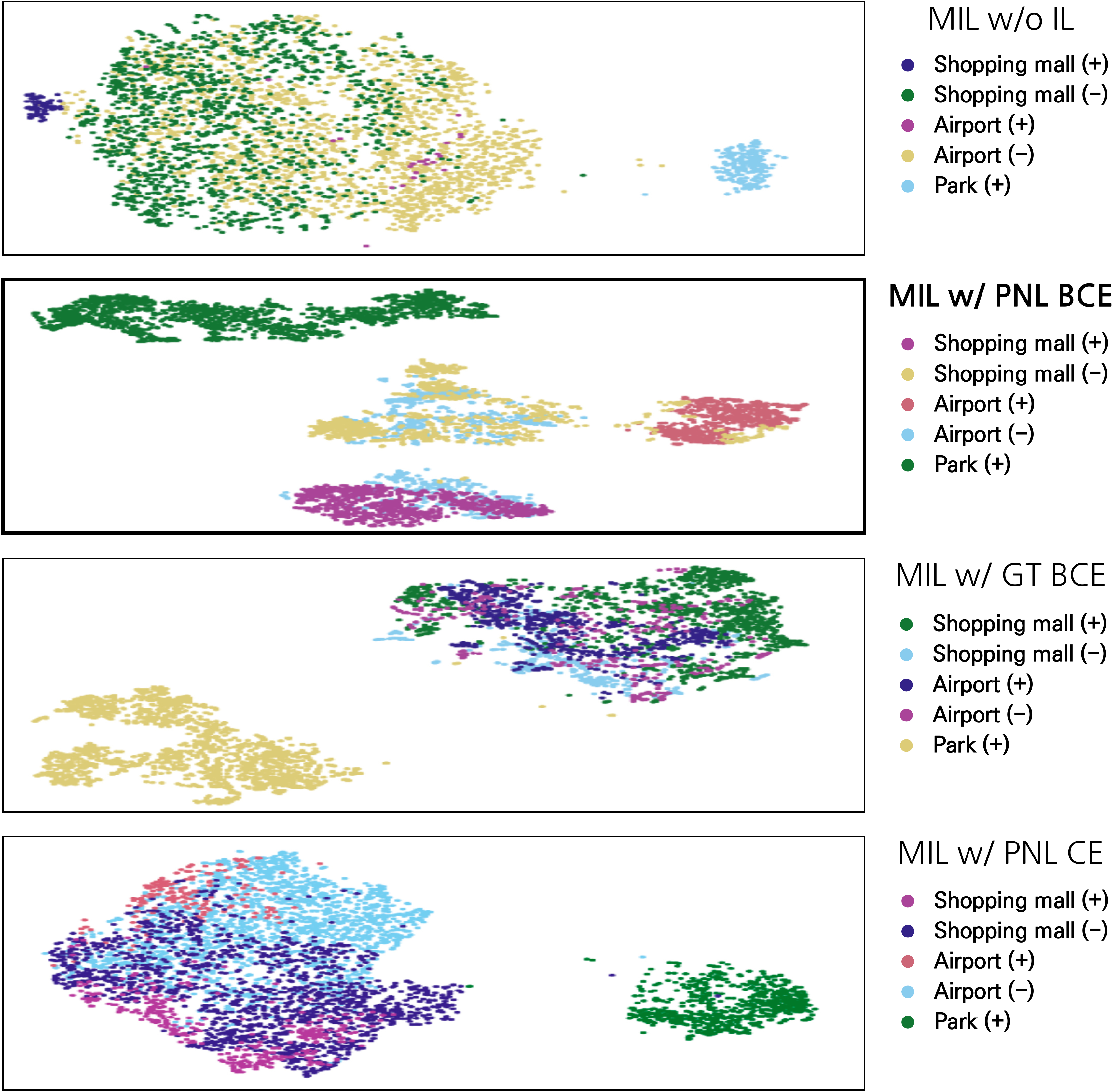}
\caption{\textbf{tSNE distribution of instances.} All points were extracted right after the instance vector generator in Fig.~\ref{fig:overall}. (\emph{Indoor}: airport, shopping mall, metro station. \emph{Outdoor}: street pedestrian, public square, street traffic, park. \emph{Transportation}: metro, bus, tram.)}
\label{fig:tsne_ambiguity}
\end{figure}

\begin{table}[h]
\centering
\caption{\textbf{Comparison of the classification performance to the frame loss.} The result is experimented on the TAU 2019.}

{
\footnotesize
\begin{tabular}{cccc} \dtoprule

\multirow{1}{*}{\makecell{Model}} & \multirow{1}{*}{\makecell{Objective function for instance-level}} & \multirow{1}{*}{Acc(\%)} \\ \midrule[1pt]
\multirow{2}{*}{\makecell{FUSE}} & \multirow{1}{*}{\makecell{Frame loss \cite{wang2022frame}}} & 71.59 \\ \cmidrule{2-3}
~ & \multirow{1}{*}{\makecell{MIL w/ PNL BCE}} & 79.13 \\ \dtoprule

\end{tabular}
}
\label{tab:frameloss}
\end{table}

As shown in Fig.~\ref{fig:TPR}, the MIL-based system training with only bag-level loss showed the underestimation problem; however, the system with instance-level loss significantly increased the confidence of instance-level predictions.
The instance-level confidence-based masked log-mel spectrograms are presented in Fig.~\ref{fig:spec}, which shows the comparison between the underestimated and well-estimated samples.
The vanilla MIL-based system made a decision the metro station audio to metro station scene, but the confidence was very low.
A system suffering the underestimation problem might result in a misclassification problem.
Meanwhile, the system with instance-level loss detected richer positive instances; the abundance of positive instances indicates that the system extracts various features from the data, resulting in improved performance.

Furthermore, instances were generated and clustered more distinctly through the backpropagation of instance-level losses.
Fig.~\ref{fig:tsne_ambiguity} shows the t-distributed stochastic neighbor embedding (tSNE) distribution of positive and negative instances in a park, shopping mall, and airport scenes.
The instances extracted by the system with instance-level loss were so well distributed similar to Fig.~\ref{fig:concept} depicted the concept of MIL.
The airport and shopping mall's acoustic scenes sound similar to an indoor category, unlike the park, which belongs to an outdoor category; label ambiguity could occur more frequently between the airport and shopping mall.
Owing to label ambiguity, as described in Fig.~\ref{fig:instance1}, the negative instances of the airport scenes were partially included in the positive instance area of the shopping mall scene, and it was the same for the negative instances of the shopping mall scenes.
However, no negative instances of either airport or shopping mall scenes were included in the park scene's positive instance area.
In this way, we can figure out the label ambiguity described conceptually in  \cite{carbonneau2018multiple}, using well-clustered instances via the proposed method.
In future work, we expect to improve the system performance by tuning based on instance distribution.

\subsubsection{MIL vs. conventional supervised learning}
\label{sec:tradSupL}
Second, the proposed method outperformed the conventional supervised learning (SupL) method.
SupL-based ASC systems aggregate a feature map to a scalar before the classifier; misclassified frames were reflected in the final inference.
Whereas, the MIL-based system could detect whether each frame composing a clip was positive for the target acoustic scene.
Then, if a positive instance was detected, the class of the acoustic scene was determined under the assumption that there were no audio clips with only negative instances.
In other words, it could be interpreted as predicting a scene using only a specific temporal section while excluding other frames that did not need to be detected within an audio clip.
As the proposed instance-level loss-based MIL framework overcame the underestimation problem, the performance outperformed the SupL-based ASC system.

\begin{figure*}[ht]
\centering
\includegraphics[width=\linewidth]{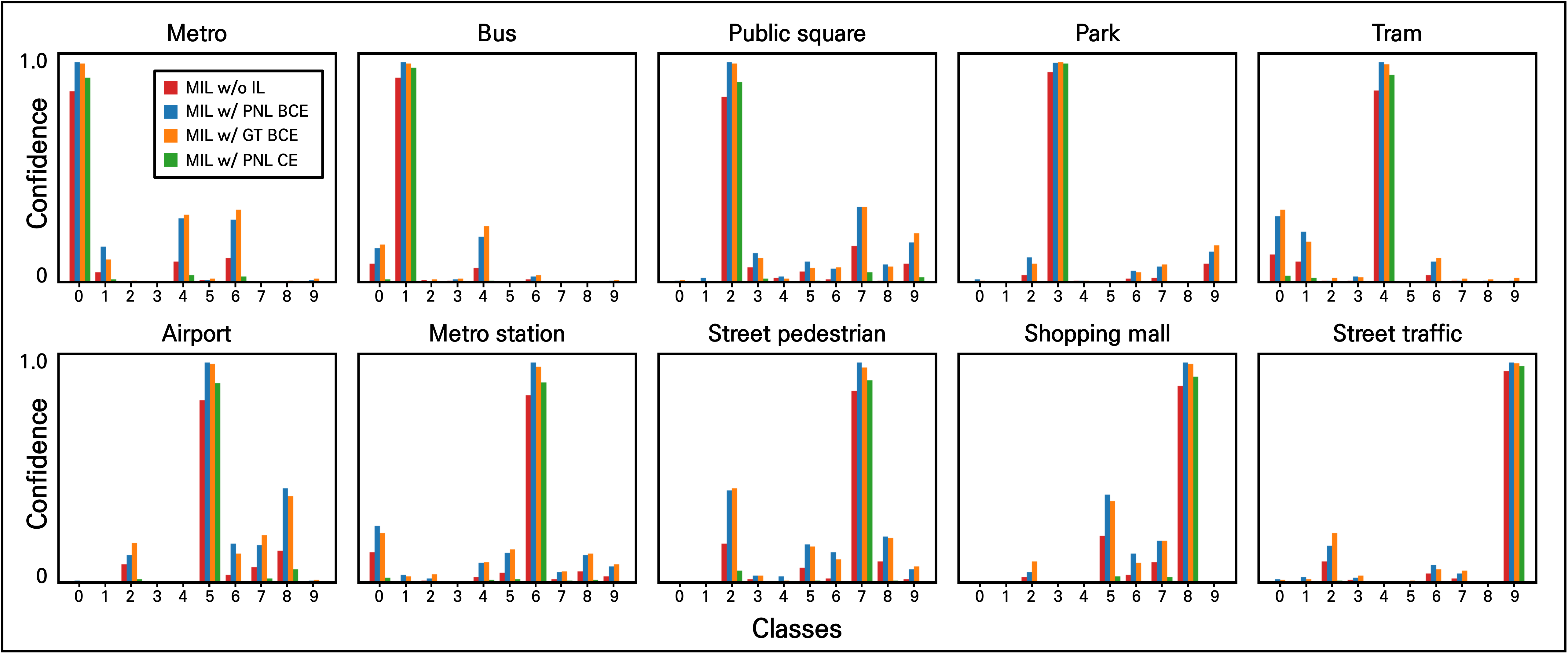}
\caption{\textbf{Averages on bag-level confidence for each class.} When there are few negative instances in the audio clip, it is difficult to distinguish the true-negative instance (ie, zero vector) from the label ambiguity with the instance-level average (Fig.~\ref{fig:TPR}). Contrary, these graphs show the bag-level confidences (i.e., the most class-wise dominant instances for each clip), and it shows how much label ambiguity is reflected when the system finally determines the scene under the SMI assumption. The difference between label ambiguity and true-negative instances could be checked from the tSNE distribution in Fig.~\ref{fig:tsne_ambiguity}. (0: metro, 1: bus, 2: public square, 3: park, 4: tram, 5: airport, 6: metro station, 7: street pedestrian, 8: shopping mall, 9: street traffic)}
\label{fig:TPR2}
\end{figure*}

\subsubsection{Bag-level ground truths for instance-level label}
\label{sec:PNvsGT}

\begin{table}[ht]
\centering
\caption{\textbf{Comparison of the classification performance according to the instance-level label: between positive/negative labels and bag-level ground truth.} FUSE was used for the feature map extractor.}

{
\footnotesize
\begin{tabular}{cccc} \dtoprule

\multirow{2}{*}{\makecell{Dataset}} & \multicolumn{2}{c}{Acc(\%)} \\ \cmidrule(lr){2-3}
 ~ & MIL w/ PNL BCE & MIL w/ GT BCE \\ \midrule[1pt]

\multirow{1}{*}{\makecell{TAU 2019}} & 79.13 & \textbf{79.45} \\ \cmidrule{1-3}
\multirow{1}{*}{\makecell{TAU 2020m}} & 66.75 & \textbf{67.94} \\ \midrule[1pt]

\end{tabular}
}
\label{tab:supL}
\end{table}

As we aforementioned at the end of Section~\ref{sec:instance_label}, we performed another experiment that assigned the bag-level ground truth to instance-level label, which combined supervised learning with the proposed MIL framework.
Table~\ref{tab:supL} and Fig.~\ref{fig:tsne_ambiguity} shows the differences in the system depending on the method for assigning instance-level label.
For every system in Table~\ref{tab:supL}, almost all performances improved slightly when the ground truth was given to instance-level label.
Furthermore, given the bag-level confidence, as shown in Fig.~\ref{fig:TPR2}, the label ambiguity seems to be considered as in the system with positive and negative labels on the instance-level label.
However, as shown in Fig.~\ref{fig:tsne_ambiguity}, the instance distribution of the third system appears confusing, except for the instances of the park, compared to the second system.
In other words, the tSNE distribution indicates that the method for assigning the ground truth to instance-level label loses its meaning as an instance of what we intended.
Although learning instances using the ground truth might be more efficient if the system is required to perform well on classification accuracy, the MIL method would be worth researching from the perspective when the instance itself is required.

\subsection{Effects of label ambiguity control: objective function}
\label{sec:objective}
It is crucial to choose an appropriate objective function based on the learning method and purpose.
When using the MIL framework for the multi-classification task, the objective function to use can be either the CE or the BCE.
As shown in Table~\ref{tab:amb}, there was no significant difference in the performance of the system in conventional supervised learning whether the loss function was the CE or the mean of the BCEs, whereas there was a significant difference when training under the MIL framework.
The experimental results show that BCE is more appropriate in MIL; in other words, we can build a more robust system when designing an MIL-based ASC system that allows label ambiguity rather than forcing it to classify ambiguous labels.

The difference is that the BCE allows label ambiguity, whereas the CE does not, because the sigmoid function evaluates the probability for each candidate, whereas the softmax function intends to choose only one target.
The details are shown in Fig.~\ref{fig:TPR2}, where the softmax score stood tall on only the target scene, but the sigmoid scores stood out even in non-target scenes that could be considered ambiguous labels.
Also, Fig.~\ref{fig:tsne_ambiguity} shows the difference in instance space according to the objective functions.
The instances were less distributed when they were forced to classify ambiguous labels compared with that using BCE.

The above results confirm that label ambiguity is natural in MIL; a bag-of-instances may have the properties of both non-target and target scenes, such as the sound of a metro can sometimes be perceived as the sound of a metro station or a tram.
The negative instance of the metro scene could fall within the metro station's concept (i.e., the positive area of the metro station scene), but it lacks significant decision-making power when inferring the bag (Fig.~\ref{fig:TPR2}).

\begin{table}[ht]
\centering
\caption{\textbf{Comparison of the classification performance according to the objective function: between BCE and CE.} The result is experimented on the TAU 2019.}

{
\footnotesize
\begin{tabular}{cccc} \dtoprule

\multirow{2}{*}{\makecell{Feature map\\extractor}} & \multirow{2}{*}{\makecell{Learning\\strategy}} & \multicolumn{2}{c}{Acc(\%)} \\ \cmidrule(lr){3-4}
~ & ~ & BCE & CE \\ \cmidrule{1-4}
\multirow{2}{*}{\makecell{FUSE}} & \multirow{1}{*}{\makecell{MIL w/ IL}}  & \textbf{78.57} & 73.41 \\ \cmidrule{2-4}
~ & \multirow{1}{*}{\makecell{SupL w/ GAP}}  & 71.59 & \textbf{71.61} \\ \midrule[1pt]

\end{tabular}
}
\label{tab:amb}
\end{table}

\begin{figure*}[b]
\centering
\includegraphics[width=\linewidth]{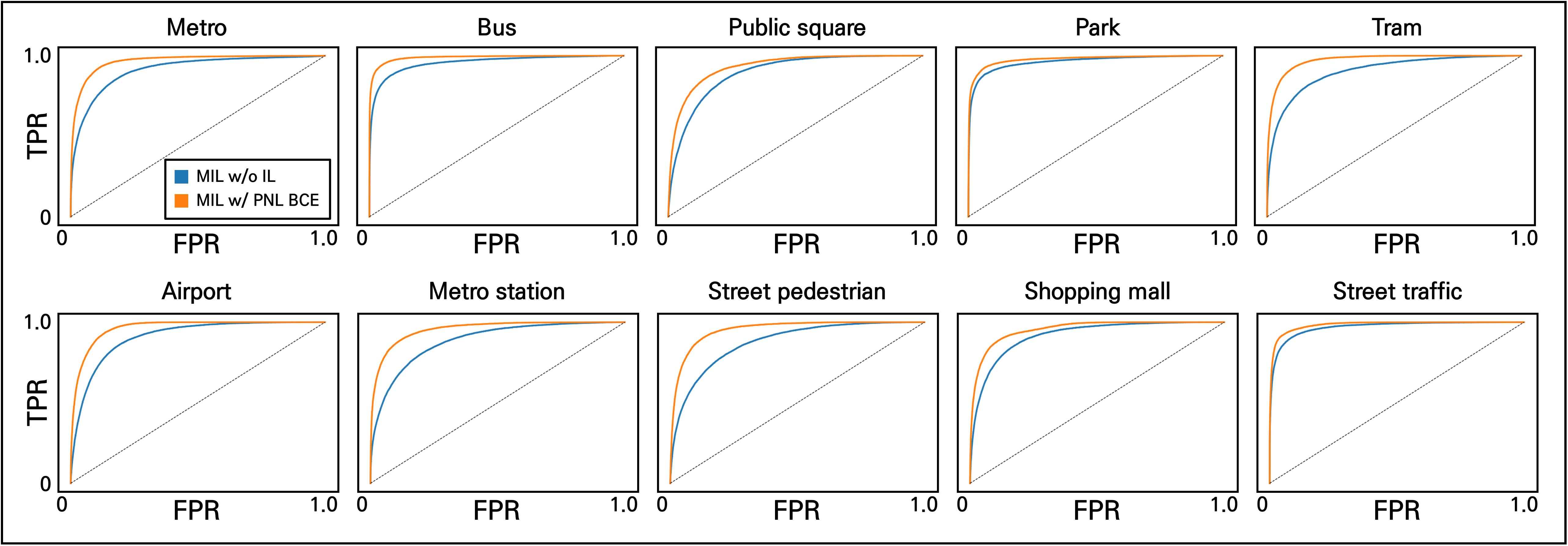}
\caption{\textbf{ROC curves for instances per each scene.} One-versus-rest strategy was used, for example, metro vs. the other scenes.}
\label{fig:ins_slosss}
\end{figure*}

\subsection{Effects of FUSE}
\label{sec:resultFUSE}
As shown in Table~\ref{tab:experiments}, the feature map extractor with FUSE had significant effects on both performance and the number of parameters compared to the VGG-like extractor in  \cite{song19b_interspeech}.
The number of parameters decayed by approximately 90\%, and the performance improved up to 6\% for each system.
Furthermore, in Table~\ref{tb:exp_result}, FUSE showed good performances under the proposed MIL framework: it outperformed the other CNNs significantly when the system was trained with a single device.
As previously mentioned, an image has rectangular-shaped local information, whereas a spectrogram has information along a harmonic or a combination of a frequency axis and continuity over a time axis.
In other words, unlike image filtering, a spectrogram plays a different role on the x- and y-axes in feature maps.
Therefore, as supported by the experimental results above, applying frequency and temporal filters separately could be more significant in feature extraction than applying a patch-like filter, particularly when training instance-wise.

\subsection{Other MIL assumptions}
\label{sec:assumption}
We used the SMI assumption for comparison to \cite{song19b_interspeech}.
For the extra study, we compared the CMI assumption to the SMI assumption.
The training strategy was same as SMI assumption but the inference was different: the bag's class was decided by the number of positive instances.
In other words, the class that had the most number of positive instance in the bag determined the bag.
Then the bag classifier \(g(\mathcal{X})\) is turned to:
\begin{equation}
\begin{split}
\hat{\mathbf{Y}} &= \sum_i^N{\mathbf{e}^{(\hat{c}_i)}}, (\hat{c}_i = \mbox{argmax}_{c}\ \hat{\mathbf{y}}_i)\\ 
g(\mathcal{X}) &= \mbox{argmax}_{c}\ \hat{\mathbf{Y}},
\label{eq:CMI}
\end{split}
\end{equation}
where \(\mathbf{e}^{(\hat{c}_i)} \in \mathbb{R}^C\) denotes one-hot vector.
The results are shown in Table~\ref{tab:assumption}.
In every case, the CMI outperformed the SMI.
In other words, there were some dominant false positive instances in some bags, although the number was small.

Also, we evaluated the proposed method using the TMI assumption.
If the TMI assumption is adopted for the multi-classification task, there could exist a singularity point where all instances are determined to be negative, so the bag could be treated as a blank scene that does not belong to any scenes.
Therefore, it is not proper to compare the classification performance to the SMI and CMI assumptions in Table~\ref{tab:amb}.
Instead, we used the receiver operating characteristic (ROC) curve for evaluation (Fig.~\ref{fig:ins_slosss}).
As the results show, the proposed method was more robust for the ASC system.

\begin{table}[h]
\centering
\caption{\textbf{Comparison of the classification performance according to the assumptions: SMI and CMI.} FUSE was used for the feature map extractor.}

{
\footnotesize
\begin{tabular}{cccc} \dtoprule

\multirow{2}{*}{\makecell{Dataset}} & \multirow{2}{*}{\makecell{Instance-level\\loss}} & \multicolumn{2}{c}{Acc(\%)} \\ \cmidrule(lr){3-4}
~ & ~ & SMI & CMI \\ \midrule[1pt]

\multirow{4}{*}{\makecell{TAU 2019}} & \multirow{1}{*}{\makecell{MIL w/o IL}} & 72.22 & \textbf{73.09} \\ \cmidrule{2-4}
~ & \makecell{MIL w/o P/N L} & 79.13 & \textbf{79.71} \\ \cmidrule{2-4}
~ & \makecell{MIL w/o  GT} & 79.45 & \textbf{80.22} \\ \cmidrule{1-4}

\multirow{4}{*}{\makecell{TAU 2020m}} & \multirow{1}{*}{\makecell{MIL w/o IL}} & 60.13 & \textbf{61.48} \\ \cmidrule{2-4}
~ & \makecell{MIL w/o P/N L} & 66.75 & \textbf{67.17} \\ \cmidrule{2-4}
~ & \makecell{MIL w/o GT} & 67.94 & \textbf{68.09} \\ \midrule[1pt]

\end{tabular}
}
\label{tab:assumption}
\end{table}

\subsection{Comparison to other systems}
\begin{table*}[ht]
\centering
\caption{\textbf{Comparisons with other systems on TAU 2019 dataset.} We recorded the best scores among the random seeds, and the CMI assumption was used. (*: Estimated number of parameters)}

{
\def\arraystretch{1.2}
\footnotesize
\begin{tabular}{lccc} \dtoprule

\multirow{1}{*}{\makecell{System}} & \multirow{1}{*}{\makecell{Augmentation}} & \multirow{1}{*}{\makecell{Params}} & \multirow{1}{*}{\makecell{Acc(\%)}} \\ \midrule[1pt]

\makecell[l]{DCASE baseline} & \makecell{-} & \makecell{116K} & 64.3 \\ \cmidrule{1-4}
\makecell[l]{Conv-StandardPOST \cite{naranjo2020acoustic}} & \makecell{-} & \makecell{528.3K} & 76.7 \\ \cmidrule{1-4}
\makecell[l]{1D-CNN based \cite{jung2020knowledge}\\ \tiny{with knowledge distillation}} & \makecell{Mixup} & \makecell{636K} & 77.6  \\ \cmidrule{1-4}
\makecell[l]{1D-CNN based \cite{jung20b_interspeech}\\ \tiny{with audio tagging}} & \makecell{Mixup \\ Specaug. \\ Slicing} & \makecell{646K} & 76.8 \\ \cmidrule{1-4}
\makecell[l]{FCNN \cite{Cho2019}} & \makecell{-} & \makecell{871K} & 76.0 \\ \cmidrule{1-4}
\makecell[l]{Deep residual net. \cite{mcdonnell2020acoustic}} & \makecell{Mixup} & \makecell{3.2M} & \textbf{81.4} \\ \cmidrule{1-4}
\makecell[l]{CNN based \cite{seo2019acoustic}} & \makecell{Mixup} & \makecell{8M} & 80.4 \\ \cmidrule{1-4}
\makecell[l]{Vgg12 \cite{Huang2019}} & \makecell{Specaug.} & \makecell{12.8M} & 77.4 \\ \cmidrule{1-4}
\makecell[l]{AlexNet-like \cite{wu2020time}} & \makecell{Mixup} & \makecell{*13M} & 76.7 \\ \cmidrule{1-4}
\makecell[l]{AclNet \cite{Huang2019}} & \makecell{Specaug.} & \makecell{19.3M} & 74.8 \\ \cmidrule{1-4}
\makecell[l]{SeNoT-Net \cite{zhang2020learning}} & \makecell{-} & \makecell{22.5M} & 80.3 \\ \cmidrule{1-4}
\makecell[l]{ATReSN-Net \cite{zhang2020atresn}} & \makecell{-} & \makecell{22.6M} & 80.7 \\ \cmidrule{1-4}
\makecell[l]{ResNet-50 \cite{Huang2019}} & \makecell{Specaug.} & \makecell{24.5M} & 77.9 \\ \cmidrule{1-4}
\makecell[l]{AclSincNet \cite{Huang2019}} & \makecell{Specaug.} & \makecell{52.2M} & 76.1 \\ \cmidrule{1-4}\morecmidrules\cmidrule{1-4}
\makecell[l]{FUSE MIL w/ PNL BCE} & \makecell{-} & \makecell{139K} & 80.3 \\ \cmidrule{1-4}
\makecell[l]{FUSE MIL w/ GT BCE} & \makecell{-} & \makecell{139K} & \textbf{81.1} \\ \midrule[1pt]

\end{tabular}
}
\label{tb:vs19}
\end{table*}

\begin{table*}[ht]
\centering
\caption{\textbf{Comparisons with other systems on TAU 2020m dataset.} We recorded the best scores among the random seeds, and the CMI assumption was used. Note that the DCASE challenge 2021 task 1-A limited the size of submitted systems, so most of the systems were built with quantization or pruning techniques for model compression. So we recorded the performance that did not use any compression techniques if there were. (*: Estimated number of parameters, **: before pruning, ***: quantized model)}

{
\def\arraystretch{1.2}
\footnotesize
\begin{tabular}{lccc} \dtoprule

\multirow{1}{*}{\makecell{System}} & \multirow{1}{*}{\makecell{Augmentation}} & \multirow{1}{*}{\makecell{Param}} & \multirow{1}{*}{\makecell{Acc(\%)}} \\ \midrule[1pt]

\makecell[l]{DCASE baseline} & \makecell{-} & \makecell{***45k} & \makecell{54.1} \\ \cmidrule{1-4}
\makecell[l]{BC-ResNet-ASC-8 \cite{Kim2021}} & \makecell{Mixup \\ Specaug. \\ Time roll} & \makecell{315K} & \textbf{75.2} \\ \cmidrule{1-4}
\makecell[l]{CP-ResNet \cite{koutini2021cpjku}} & \makecell{-} & \makecell{65K \\ **635K} & 69.5  \\ \cmidrule{1-4}
\makecell[l]{LIC-TSL \cite{Yang2021}} & \makecell{-} & \makecell{*882K} & 69.9  \\ \cmidrule{1-4}
\makecell[l]{GATSCNN-2 \cite{kek2022multi}} & \makecell{Mixup} & \makecell{*4.87M} & 73.3 \\ \cmidrule{1-4}
\makecell[l]{Three-stream ResNet \cite{jo2021global}} & \makecell{Mixup} & \makecell{6.42M} & 72.5 \\ \cmidrule{1-4}
\makecell[l]{DcaseNet-v3 \cite{jung2021dcasenet}} & \makecell{Mixup} & \makecell{13.2M} & 70.35 \\ \cmidrule{1-4}
\makecell[l]{Trident ResNet \cite{suh2020designing}} & \makecell{Mixup \\ Temporal crop} & \makecell{13M} & \makecell{73.7} \\ \cmidrule{1-4}\morecmidrules\cmidrule{1-4}
\makecell[l]{FUSE MIL w/ PNL BCE} & \makecell{-} & \makecell{139K} & 67.5 \\ \cmidrule{1-4}
\makecell[l]{FUSE MIL w/ GT BCE} & \makecell{-} & \makecell{139K} & 69.4 \\ \cmidrule{1-4}
\makecell[l]{FUSE MIL w/ GT BCE\\ \tiny{with pre-ResNorm}} & \makecell{-} & \makecell{139K} & \textbf{72.3} \\ \midrule[2pt]
\end{tabular}
}
\label{tb:vs20}
\end{table*}

To validate the contributions of the proposed method, we compared it with recent ASC studies and systems in DCASE challenges that used single networks without any ensemble techniques.
The performance of systems without data augmentation was reported if there was.

The results show the effectiveness of our proposed system in terms of both classification accuracy and number of parameters.
On the TAU 2019 dataset, our systems had the fewest parameters except for the DCASE baseline, and \emph{FUSE MIL-GT} had the best classification performance among the systems with less than 1 M parameters.
Furthermore, the proposed method outperformed systems with more than 1 M parameters except for the deep residual network  \cite{mcdonnell2020acoustic}.

Contrary to the ASC performance under a single device, the proposed system requires more research on multi-device ASC (i.e., device generalization for minor and unseen devices).
On the TAU 2020m dataset, our systems also had the fewest parameters except for the DCASE baseline, but performed poorly in comparison to the other systems.
However, when using residual normalization (ResNorm), which is known to be good for device generalization  \cite{Kim2021}, the accuracy outperformed other systems.
Although the performance improved by approximately 2.5\% by applying ResNorm, further research is required to maintain the scene classification performance in unseen devices.

Fortunately, we showed that assigning positive and negative labels for instance-level labels was good for instance clustering in this study.
Therefore, for future work, we plan to improve the proposed system into an MIL-based recording device-invariant ASC system with similar model complexity, in which instances will be embedded in the instance-space regardless of device characteristics.
In the following paragraphs, we concisely summarize some systems that outperformed our proposed method or had a similar motivation.

\textbf{Deep residual network} \cite{mcdonnell2020acoustic} had two streams based on frequency bands.
Each stream consisted of a different CNN: one was for the low-frequency band (i.e., the half-low of the frequency bins) and the other for the high-frequency band.
The network structure's motivation was similar to that of FUSE: spectrograms have different characteristics to images, non-locality, and different properties for different frequency bands.
However, they approached it by passing it through different networks for different frequency bands at an early stage and performed late fusion.

\textbf{1D-CNN based model} \cite{jung2020knowledge} shows the best performance in Table~\ref{tb:vs19} with under 1 M parameters, except for our proposed system.
Unlike most other systems that use the log-mel spectrogram for the input data, this system used a raw wave signal and consisted of 1D convolutional layers with a support vector machine (SVM) as a classifier.
Furthermore, the system was trained with a knowledge distillation strategy, in which the teacher model generated soft labels for the student model with a longer audio duration (i.e., 20 s or 30 s consisting of multiple audio segments).
The longer the recording time, the more various instances of the acoustic scene can be included, allowing the teacher model to generate more sophisticated and generic labels.

\textbf{BC-ResNet-ASC-8} \cite{Kim2021} was the winner system in the DCASE challenge 2021 task 1-A.
BC-ResNet was first proposed for a low-complexity keyword spotting system that consisted of FDW and TDW convolutional layers to reduce model complexity in each stage \cite{kim2021broadcasted}.
In contrast  to FUSE, TDW convolutional layers were applied after averaging the feature map by the frequency axis; the result was then summed up by expanding by the frequency axis to the identity shortcut.
Furthermore, according to a technical report  \cite{kimqti}, one of the significant factors contributing to the performance under the multi-device was ResNorm, which increased the classification accuracy by approximately 6\% more than without it.
The motivation for ResNorm was batch-instance normalization  \cite{nam2018batch}, which had a significant impact on the style-invariant system in image style transfer task by combining batch-normalized features and instance-normalized features.

\textbf{Trident ResNet}  \cite{suh2020designing} was the winner system in the DCASE challenge 2020 task 1-A.
There were two main strategies for improving the system under recording with different device characteristics.
First, it was split into three streams by the frequency axis, similar to the \emph{deep residual network}  \cite{mcdonnell2020acoustic}, which was found through grid searching for the number of streams to be divided into for device generalization.
The other was that the focal loss \cite{lin2017focal}, first proposed for the imbalanced class in a dataset, which was adopted for the objective function, and Trident ResNet used it to alleviate the imbalanced-device problem.
Additionally, it used pre-activation  \cite{he2016identity}, a CNN with a normalization-activation function-convolutional layer order, and the deltas and delta-deltas features of the log-mel spectrogram as the input of the network as well as the log-mel spectrogram.
Trident ResNet performed well when it had large parameters, but had minor effects when it was turned into a lightweight system despite considering the application of quantization \cite{Jeong2021}.

\textbf{Three-stream ResNet} \cite{jo2021global} had almost the same architecture as \emph{Trident ResNet}, but it was a different version.
In this study, researchers showed that the multi-stream network for ASC performed best performance when the streams were constructed with a low-frequency stream, high-frequency stream, and global stream (i.e., two streams were for the split spectrogram, and the other was for the non-split spectrogram), allowing the network to capture global-local information from a log-mel spectrogram.
Note that the number of parameters was approximately half that of the \emph{Trident ResNet}; therefore, the comparison only in Table~\ref{tb:vs20} may cause confusion.

\section{Conclusions}

In this study, we propose an improved MIL framework that is effective in extracting instances.
The proposed MIL framework defines instance-level labels using bag-level labels, which are used when calculating instance-level loss.
Assigning positive and negative labels for instance-level loss allows the system to detect positive instances more accurately and describe the relationship between the data.
The abundance of positive instances indicates that the system is effective in extracting a wide range of distinct features from the data.
Furthermore, we proposed a novel architecture for the instance generator FUSE that only consists of column-like convolutional filters, and dramatically reduces the number of parameters.
The experimental results show that the proposed MIL framework estimated more positive instances than the vanilla MIL-based ASC system.
The proposed system exhibited remarkable performance with a lower number of parameters, particularly on single-device recordings.
Finally, we anticipate that the method for assigning positive and negative labels for the instance-level loss could be widely applied to other MIL tasks as well as the ASC.

\clearpage
\bibliographystyle{IEEEtran.bst}
\bibliography{refs.bib}

\end{document}